\documentclass[12pt]{book}

\usepackage[dvips]{graphicx,color}
\usepackage{makeidx,tsukuba}

\makeauthorindex
\makeindex

\begin{document}

\BookTitle{\itshape The 28th International Cosmic Ray Conference}
\CopyRight{\copyright 2003 by Universal Academy Press, Inc.}
\pagenumbering{arabic}

\chapter{A Fast Hybrid Approach to Air Shower Simulations and Applications}

\author{%
Hans-Joachim Drescher$^1$, Glennys Farrar$^2$, Marcus Bleicher$^1$, 
 Manuel Reiter$^1$, Sven Soff$^1$, Horst St\"ocker$^1$ \\
{\it (1) Institut f\"ur Theoretische Physik, Johann Wolfgang Goethe Universit\"at, Robert-Mayer Str. 8-10, Frankfurt am Main, Germany\\
(2) Center for Cosmologie and Particle Physics, New York University, 4 Washington Place, NY 10003, USA}\\
}

\section*{Abstract}
The SENECA model, a new hybrid approach to air shower simulations, 
is presented.
It combines the use of efficient cascade equations in the energy 
range where a shower can be treated as one-dimensional, with a traditional
Monte Carlo method which traces individual particles.
This allows one to reproduce natural fluctuations of individual 
showers as well as the lateral spread of low energy particles. 
The model is quite efficient in computation time.

As an application of the new approach, the influence of the low energy hadronic
models on shower properties for AUGER energies is studied. 
We conclude that these models have a significant impact on the tails 
of lateral distribution functions, and deserve therefore more attention. 

\section{Introduction}

In the field of air shower simulations, large computation times limit 
the applicability of traditional Monte-Carlo approaches 
at very high energies. The thinning algorithm tries solve this 
problem, at the price of introducing artificial fluctuations. 
Here, we present a hybrid approach to air shower simulation, 
as introduced in [2]: A system 
of transport equations is employed in the energetic region where an air
shower can be considered as one dimensional. The lateral spread of low 
energy particles is calculated with traditional Monte-Carlo methods,
where one traces each particle individually. 

\section{Hadronic Cascade Equations}

If one considers an air shower as a one-dimensional system, it can 
be described with the following set of cascade equations (CE) [1,2]: 

\begin{eqnarray}
\frac{\partial h_{n}(E,X)}{\partial X} & = & -h_{n}(E,X)\left[ \frac{1}{\lambda _{n}(E)}+\frac{B_{n}}{EX}\right] \label{for:hce} \\
 &  & +\sum _{m}\int _{E}^{E^{\mathrm{had}}_{\mathrm{max}}}h_{m}(E',X)\left[ \frac{W_{mn}(E',E)}{\lambda _{m}(E')}\right. \nonumber 
\left. +\frac{B_{m}D_{mn}(E',E)}{E'X}\right] dE'\nonumber ~,
\end{eqnarray}

where $h_{n}(E,X)dE$ is the number of particles of type $n$ (basically nucleons,
pions and kaons) with an energy $[E,E+dE]$ at an atmospheric depth $X$. 
The first line with the minus-sign accounts for particles falling out
of the system by either inelastic collisions with air-nuclei or by decay. 
The second line accounts for producing particles by the same two mechanisms
at higher energies, $W_{mn}(E,E') = \frac{dN}{dE'}$ is the differential 
energy spectrum for collisions of particle type $m$ at energy $E$
 and $D_{mn}$ are the corresponding decay functions. The technique to 
solve these equations is described  in detail in references [1,2]. 

\section{Electromagnetic Cascading} 

Electromagnetic cascades can in principle be described in a similar way. 
However, due to the fact that no decays are involved it is sufficient 
to compute the energy spectra of shower particles 
for a reference layer of air of a given thickness, e.g. $2.5~{\rm g/cm^2}$. 
We use discretized energies as $E_i=10^{(i-1)/n_{d}}~{/rm GeV}$, with
$n_d$ bins per decade.  
If $V^{mn}_{ji}$ describes the energy of electrons/positrons and photons, 
for a particle of type $m$ traversing this layer of air, then 
a simple iteration

\begin{equation}
g^{n}_{i}(X+\Delta X)=\sum _{m,j \ge i}g^{m}_{j}(X)V^{mn}_{ji}(\Delta
X)\, .
\end{equation}
describes the shower at any depth $X$. 

\section{Source Functions}

The system of hadronic and electromagnetic cascade equations is solved down 
to energies of $1000$~GeV and $10$~GeV, respectively. Below these energies,
the three-dimensional character of the air shower becomes important. 
Thus, instead of evolving further, we calculate directly the 
energy spectra of low energy particles in the following way:

\begin{eqnarray}
\frac{\partial h^{\rm source}_{n}(E,X)}{\partial X} & = & \sum _{m}\int _{E^{\mathrm{had}}_{\mathrm{min}}}^{E^{\mathrm{had}}_{\mathrm{max}}}h_{m}(E',X)\left[ \frac{W_{mn}(E',E)}{\lambda _{m}(E')}\right. \nonumber
+\left. \frac{B_{m}D_{mn}(E',E)}{E'X}\right] {\rm d}E'\,. \label{for:hsource} 
\end{eqnarray}
 The source function is then used to generate low energy particles which are 
then followed in a standard Monte-Carlo approach. 

\section{Tests of Results}

\begin{figure}[t]
  \begin{center}
    \includegraphics[width=7cm]{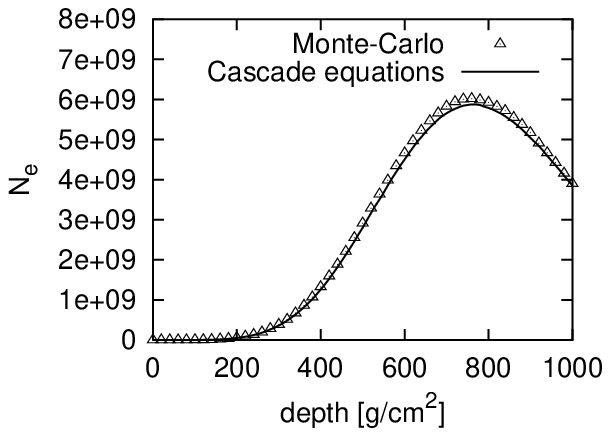}
    \includegraphics[width=7cm]{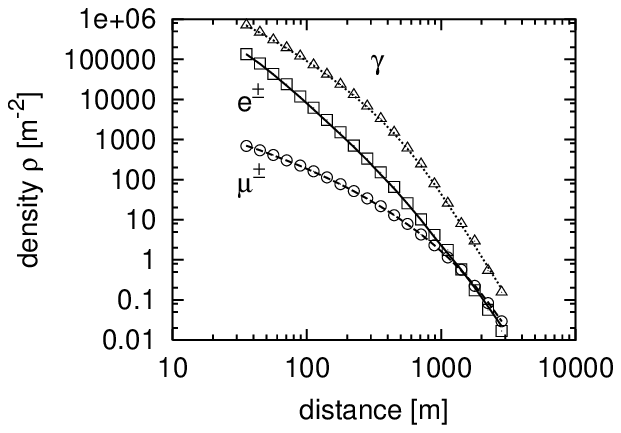}
    \includegraphics[width=7cm]{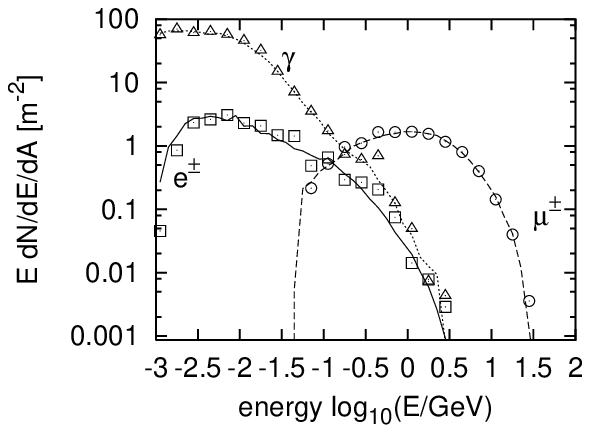}
    \includegraphics[width=7cm]{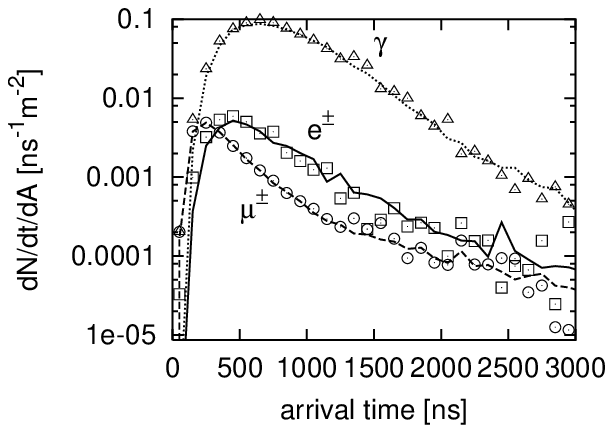}
  \end{center}
  \caption{Comparison of the hybrid method (lines) with the traditional Monte-Carlo (symbols).}
\end{figure}

The use of cascade equations should be of technical nature only and not 
change the physics results. Therefore we compare results to the 
pure Monte Carlo 
approach with thinning. The average of 500 proton induced $10^{19}$~eV
showers with thinning level $10^{-5}$ has been computed. Fig. 1 shows 
comparisons for longitudinal and lateral profiles as well as the 
energy and arrival time spectra at 1000~m distance from the shower axis. 
The agreement between the CE and the MC approach is satisfactory.

\section{Low Energy Model Dependence on Lateral Distribution Functions}

\begin{figure}[t]
  \begin{center}
    \includegraphics[width=7cm]{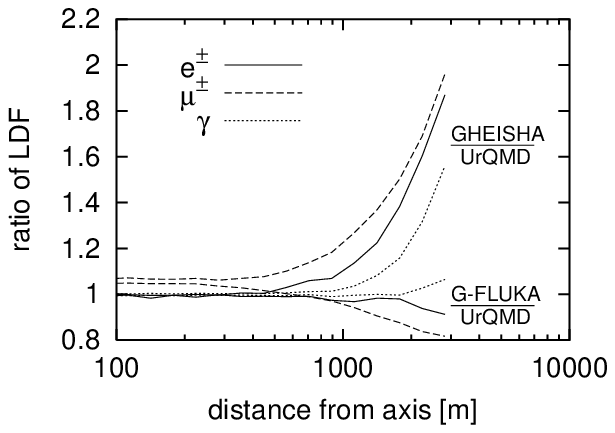}
    \includegraphics[width=7cm]{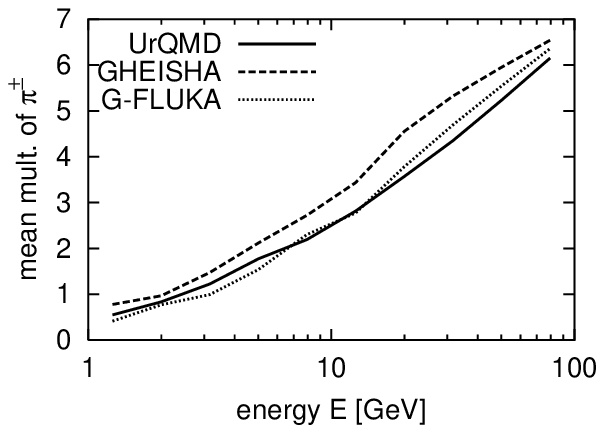}
  \end{center}
  \caption{Left:Ratio of LDF of G-FLUKA and GHEISHA to UrQMD as low energy hadronic model. Right: yield of $\pi^{+-}$ production
as a function of energy.}
\end{figure}

In Ref. [3] it was pointed out that the low-energy hadronic 
model is of crucial importance for muons in general, and for the tails 
the lateral distribution functions (LDF)
 of electrons and photons. The reason was that electromagnetic 
subshowers contributing to large distances are emitted by low 
energy particles at a large distance from the shower axis. 

Here we want to analyse the influence of different low energy 
hadronic models on the LDF of electrons, muons and photons. 
We use the following models: GHEISHA [7] and G-FLUKA [8] as found 
in GEANT3.21, and the UrQMD [5,6] model developed at the University of 
Frankfurt. For high energies we employ the QGSJET01 [4] model. 
In Fig. 2 we show the 
ratio of LDFs of $e^{\pm}, \mu^{\pm}$ and $\gamma$ 
for the different models. We see how the tails of the 
distributions change due to different choices of the 
low-energy hadronic model. GHEISHA produces more muons than UrQMD, 
since it gives a larger yield for charged pion production 
as seen in Fig. 2.
The differences even influence the tails of the LDF of electrons 
and photons, though less pronounced.
The results for UrQMD and G-FLUKA are more similar to each other.

\section{Acknowledgments}
This works has been supported by NASA-Grant NAG-9246 and by the German BMBF/Desy. The computational resources were provided by the Center for Scientific 
Computing Frankfurt (CSC).
\section{References}

\re
1.\ Bossard et al.,Phys.Rev.D63:054030,2001.
\re
2.\ H.J. Drescher, G.R. Farrar, [astro-ph/0212018], in print Phys.Rev.D 
\re
3.\ H.J. Drescher, G.R. Farrar, Astropart.Phys.19:235-244,2003
\re
4.\ N.N. Kalmykov and S.S. Ostapchenko and A.I. Pavlov ,Nucl. Phys. B 17,1997
\re 
5.\ S.A.Bass et al. Prog.Part.Nucl.Phys. 41 (1998) 225 
\re
6.\ M.Bleicher et al. J.Phys. G25  (1999) 1859
\re
7.\ H.Fesefeldt PITHA 85/02, Aachen 1985
\re 
8.\ A.Fasso et al. Proceedings of the Workshop on
Simulating Accelerator Radiation Environments, Santa Fe, 1993

\endofpaper
\end{document}